\documentclass[traditabstract]{aa}
\usepackage{natbib}
\usepackage{epsfig}
\usepackage{txfonts}
\usepackage{graphicx}

\def\ms{\,m\,s$^{-1}$}         %m.s -1
\def\kms{\,km\,s$^{-1}$}         %m.s -1
\def\vsini{$v$\,sin\,$i$}      %vsini
\def\ms{\hbox{\,m\,s$^{-1}$}}         %m.s -1
\def\m2s2{\hbox{\,m$^{2}$\,s$^{-2}$}} %m2.s -2
\def\kms{\hbox{\,km\,s$^{-1}$}}       %km.s -1
\def\vsini{\hbox{$\upsilon \sin i$}}      %vsini
\def\Msun{\hbox{$\mathrm{M}_{\odot}~$}}             %Msun
\def\Rsun{\hbox{$\mathrm{R}_{\odot}~$}}
\def\Mjup{\hbox{$\mathrm{M}_{\rm Jup}$~}}
\def\Rjup{\hbox{$\mathrm{R}_{\rm Jup}$~}}

\def\teff{T$_{\rm eff}~$}
\def\logg{log {\it g}}
\def\met{[Fe/H]~}
\def\mr{$M_\star^{1/3}/R_\star$}

\begin{document}
\title{SOPHIE velocimetry of \textit{Kepler} transit candidates \thanks{Based on observations made with the 1.93-m telescope at Observatoire de Haute-Provence (CNRS), France}}
\subtitle{II. KOI-428b: a hot Jupiter transiting a subgiant F-star}

\author{
 A. Santerne \inst{1, 3} 
\and R.~F. D\'iaz \inst{2,3}
\and F. Bouchy \inst{2, 3}
\and M. Deleuil \inst{1}
\and C. Moutou \inst{1}
\and G. H\'ebrard \inst{2,3}
\and A. Eggenberger \inst{4}
\and D. Ehrenreich \inst{4}
\and C. Gry \inst{1}
\and S. Udry \inst{5}
}

\institute{
Laboratoire d'Astrophysique de Marseille, Universit\'e d'Aix-Marseille \& CNRS, 38 rue Fr\'ed\'eric Joliot-Curie, 13388 Marseille cedex 13, France
\and Institut d'Astrophysique de Paris, UMR7095 CNRS, Universit\'e Pierre \& Marie Curie, 98bis boulevard Arago, 75014 Paris, France
\and Observatoire de Haute Provence, Universit\'e d'Aix-Marseille \& CNRS, 04670 Saint Michel l'Observatoire, France
\and Laboratoire d'Astrophysique de Grenoble, Universit\'e Joseph Fourier, CNRS (UMR 5571), BP 53, Grenoble cedex 9, France
\and Observatoire de Gen\`eve, Universit\'e de Gen\`eve, 51 Ch. des Maillettes, 1290 Sauverny, Switzerland
}
\date{Received 15 September 2010; Accepted 31 December 2010}

\offprints{A.~Santerne\\
 \email{alexandre.santerne@oamp.fr}}

\abstract{ We report the discovery of a hot Jupiter transiting a subgiant star with an orbital period of 6.87 days thanks to public photometric data from the \textit{Kepler} space mission and new radial velocity observations obtained by the SOPHIE spectrograph. The planet KOI-428b with a radius of $1.17 \pm 0.04 $ \Rjup and a mass of $2.2 \pm 0.4$  \Mjup, orbits around a F5IV star with $R_{\star}$ = 2.13 $\pm$ 0.06 \Rsun, $M_\star$= $1.48 \pm 0.06 $ \Msun ~ and \teff= 6510 $\pm$ 100 K. The star KOI-428 is the largest and the most evolved star discovered so far with a transiting planet.

\keywords{planetary systems -- techniques: photometry -- techniques:
  radial velocities - techniques: spectroscopic, star : individual(KOI-428, KIC10418224, 2MASS 19471528+4731357) }
}

%\titlerunning{\textit{SOPHIE} follow-up of \textit{Kepler} planetary transiting candidates I.}
\titlerunning{KOI-428b, a hot Jupiter transiting a subgiant F-star}
\authorrunning{A.~Santerne}

\maketitle

\section{Introduction}
\label{intro}
Launched in march 2009, \textit{Kepler} is the second space mission designed to find transiting exoplanets with high accuracy photometry. This mission has already shown its capability to discover new planets during the first 33.5 days of science operations with the announcement of a hot-Neptune-like planet \citep{2010ApJ...713L.126B}, four hot-Jupiters \citep{2010ApJ...713L.131K, 2010arXiv1001.0416J, 2010ApJ...713L.136D, 2010ApJ...713L.140L}, and a system of two transiting Saturns \citep{holmanetal}. From the 156,000 stars observed, 706 exoplanet candidates were identified and 306 of them were published by \citet{2010arXiv1006.2799B}. All these candidates are detected around faint stars (14 $\le$ mV $\le$ 16) not included in the ground-based follow-up conducted by the Kepler team. 

To establish the planetary nature of these candidates and to assess the fraction of false positives, a {high-resolution spectroscopic follow-up must be carried out. This could allow various configurations of eclipsing binaries to be discarded and the properties (mass, density) of actual planets to be established \citep{2009IAUS..253..129B}. We present here the radial velocity follow-up performed with the SOPHIE spectrograph, mounted on the 1.93-m telescope at Observatoire de Haute Provence, of the Kepler Object of Interest KOI-428 as part of a first set of \textit{Kepler} targets followed-up with SOPHIE (Bouchy et al., in prep). This led to the discovery and the characterization of the new exoplanet KOI-428b.

\section{\textit{Kepler} observations  \label{sect.keplerdata}}

The target KOI-428 was observed by \textit{Kepler} during the first 33.5-day segment of the science operations (Q1) from May, 13 to June, 15, 2009 with a temporal sampling of 29.4 min. The various identifications of this target, including coordinates and magnitudes are listed in Table \ref{keplerID}. The publicly-available\footnote{\url{http://archive.stsci.edu/kepler/data_search/search.php}} \textit{Kepler} light curve contains1639 points, thirteen of which are affected by known systematic effects, such as the loss of fine pointing caused by momentum desaturation or 
the occurrence of strong argabrightenings\footnote{Name given to cadences for which all the focal plane is illuminated by a presently unexplained effect. For more informations, see \url{http://archive.stsci.edu/kepler/release\_notes/release\_notes5/Data\_Release\_05\_2010060414.pdf}, Section 6.1}, and were discarded by the \emph{Kepler} pipeline \citep{2010ApJ...713L..87J}. Based on the four transits that occurred during Q1, \citet{2010arXiv1006.2799B} report that this system consists of a $\sim$1.93 solar radius star with a $1.04$ Jupiter radius companion in a 6.87-days orbit.  Figure \ref{428fullLC} presents the corrected light curve from the MAST database showing four periodic transit events with a depth of about 3.9 mmag. One can see from this figure that the first event is slightly deeper than the other three. We therefore studied if other known systematic effects were present during the transits, in particular during the first one. We found that none of the transits were affected by Reaction Wheel Zero Crossings or by a fainter argabrightenings, which were not discarded by the pipeline. Also, no difference was observed in the out-of-transit flux of the first transit as compared to the other three, which would have indicated that the difference in depth of the first transit was caused by the detrending performed by the pipeline. On the other hand, it is known that during Q0 and Q1 a couple of variable stars --an eclipsing binary and an intrinsic variable-- were chosen as guide stars, and that this affected the spacecraft guiding. As a consequence, the time-series of the column position of KOI428 presents positive sharp transit-like features lasting about 10-15 cadences. One of these "bumps" coincides with the first transit and might explain the amplitude difference observed. 
%This may be due to an imperfect correction of systematic effects coming from the image shift on the CCD. 
Since the inclusion of this transit in our fitting procedure does not change our results significantly, we decided to use the whole \emph{Kepler} light curve as is provided in the MAST database.

\begin{table}[]
\centering
\caption{KOI\-428 IDs, coordinates and magnitudes}
\renewcommand{\footnoterule}{}                          
\begin{minipage}[t]{7.0cm} 
\begin{tabular}{cc}
\hline
\hline
Kepler Input Catalog (KIC) & 10418224 \\
Kepler Object of Interest (KOI) & 428.01 \\
2MASS ID & 19471528+4731357 \\
%GSC2.3 & N2HG047856\\
%USNO-A2 & 1350-11065312\\
 \hline
Right Ascension (J2000) & 19:47:15.29 \\
Declinaison (J2000) & +47:31:35.8 \\
\hline
Kepler magnitude$^{(a)}$ & 14.58 $\pm$ 0.02\\
 $B^{(b)}$   & 14.9\\ 
 $R^{(b)}$  & 14.3 \\
 $J^{(c)}$ & 13.536 $\pm$ 0.021\\
 $H^{(c)}$ & 13.294 $\pm$ 0.022\\
 $K^{(c)}$ & 13.27 $\pm$ 0.30\\
 $E(B-V)^{(a)}$ & 0.139 $\pm$ 0.1\\
\hline
\hline
\end{tabular}
%\footnotetext[1]{from the Tycho-2 catalog, the Kepler magnitude ($K_p$) is calculated as follows : $blue = 0.54*B + 0.46*V - 0.07$, $red = -0.44*B + 1.44*V + 0.12$, $color = blue - red$. If $color \leq 0.8$ $K_p = 0.8 * red + 0.2 * blue$ else $K_p = 0.9 * red + 0.1 * blue$}
\vspace{-0.3cm}
\footnotetext{$^{(a)}$ from the Kepler Input Catalog.~$^{(b)}$ from USNO-A2 catalog.~$^{(c)}$ from 2MASS catalog.}
\label{keplerID}      
\end{minipage}
\end{table}

\begin{figure}[]
\begin{center}
\includegraphics[width=8.5cm]{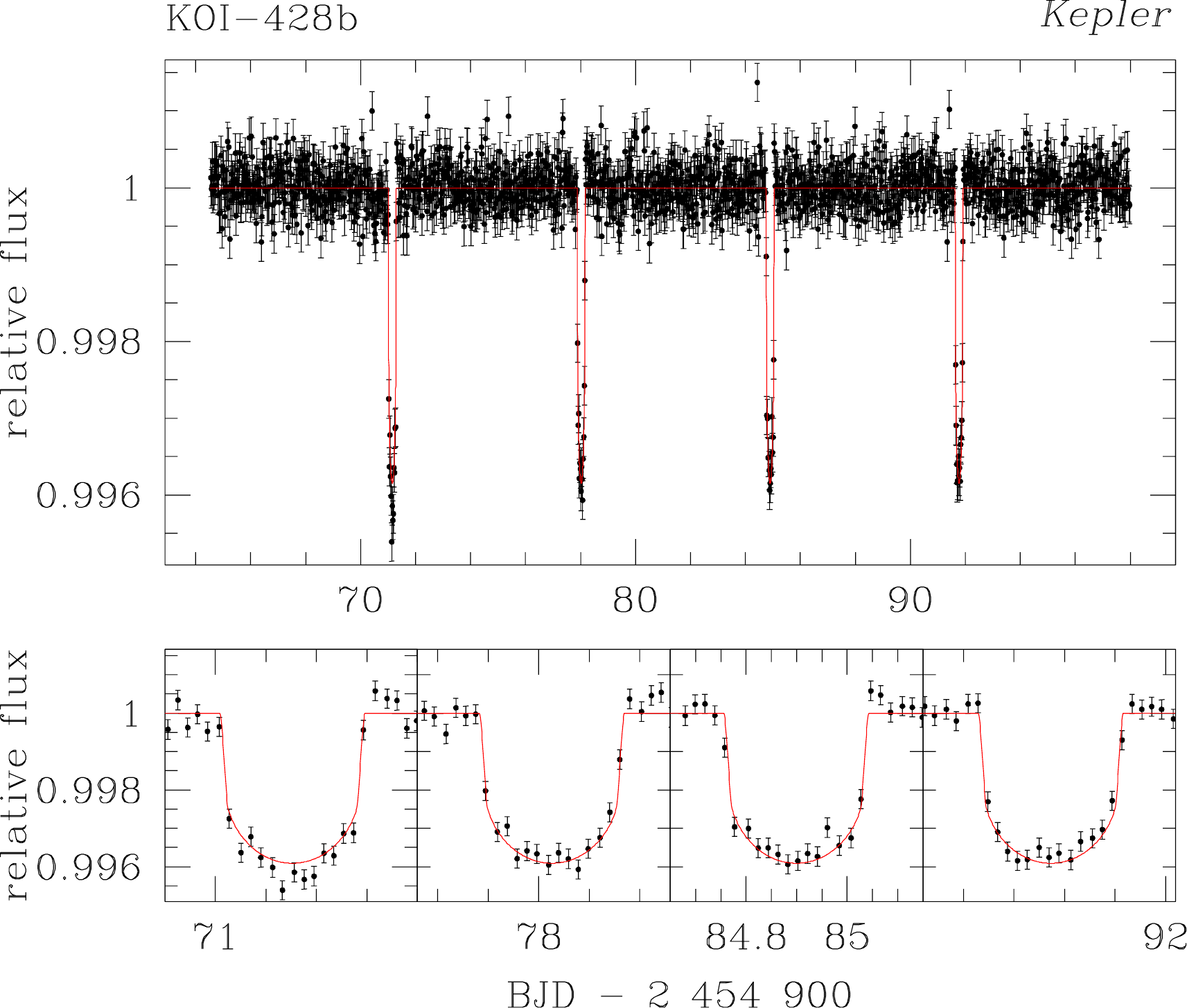}
\caption{Light curve available on MAST database showing the four transits detected on KOI-428. The red curve shows the best fit of the transit.}
\label{428fullLC}
%\vspace{-0.5cm}
\end{center}
\end{figure}

\section{SOPHIE observations}

We performed the high-resolution spectroscopy follow-up of KOI-428 as part of a first set of \textit{Kepler} candidates (Bouchy et al. in prep.) with the SOPHIE spectrograph \citep{2008SPIE.7014E..17P, 2009A&A...505..853B} installed on the 1.93m telescope at the Observatoire de Haute-Provence, France. SOPHIE is a cross-dispersed high-resolution fiber-fed echelle spectrograph, stabilized in pressure and temperature and calibrated with a Thorium-Argon lamp, mainly dedicated to measure precise radial velocity (RV) on solar-type stars for exoplanets and asteroseismology studies. 

Observations were made with the high efficiency mode, with a spectral resolution of 39,000 at 550 nm, and the slow read-out CCD mode. Our observations made use of both of the SOPHIE fibers : fiber A was centered on the target, while fiber B acquired the sky background $\sim$ 2 arcmin from the target. The typical intrinsic stability of SOPHIE (3 \ms h$^{-1}$) does not require to use the simultaneous calibration and the sky fiber is critical to remove the scattered Moon light. Eleven spectra of KOI-428 were obtained from July, 15 to September, 13, 2010 with an exposure time of 1 hour.  SOPHIE spectra were reduced with the online pipeline. Radial velocities were obtained by computing the weighted cross-correlation function (CCF) of the spectra using a numerical spectral mask corresponding to a G2V star \citep{1996A&AS..119..373B, 2002A&A...388..632P}. 
Four spectra were strongly affected by the Moon background light and were corrected with the same procedure as for previous planets \citep{2008MNRAS.385.1576P, 2008A&A...482L..17B, 2008A&A...488..763H}\textbf{,} and as described for \textit{HARPS} \citep{2010arXiv1006.2949B}. This effect cannot be removed completely on such a faint star, hence we conservatively doubled the uncertainties of the four corresponding measurements to account for possibly remaining systematics and to get a reduced $\chi^2$ close to one. Radial velocities are listed in Table \ref{428rv} and plotted in Fig. \ref{428ph}. They show a clear variation in phase with the \textit{Kepler} ephemeris listed in Table \ref{starplanetparam} and compatible with the reflex motion of the parent star due to a planetary companion. Since no significant eccentricity is seen in the data (e = $0.05 \pm 0.21$) we decide to fit a Keplerian circular orbit using the updated ephemeris listed in Table \ref{starplanetparam}. This does not affect the radial velocity semi-amplitude within 1-$\sigma$ level. The best-fit has a semi-amplitude $K$ = 179 $\pm$ 27 \ms~ and a $\sigma_{O-C} = 65 \ms$ which is comparable to the mean radial velocity uncertainty.

\begin{figure}[]
\begin{center}
\includegraphics[width=8.5cm]{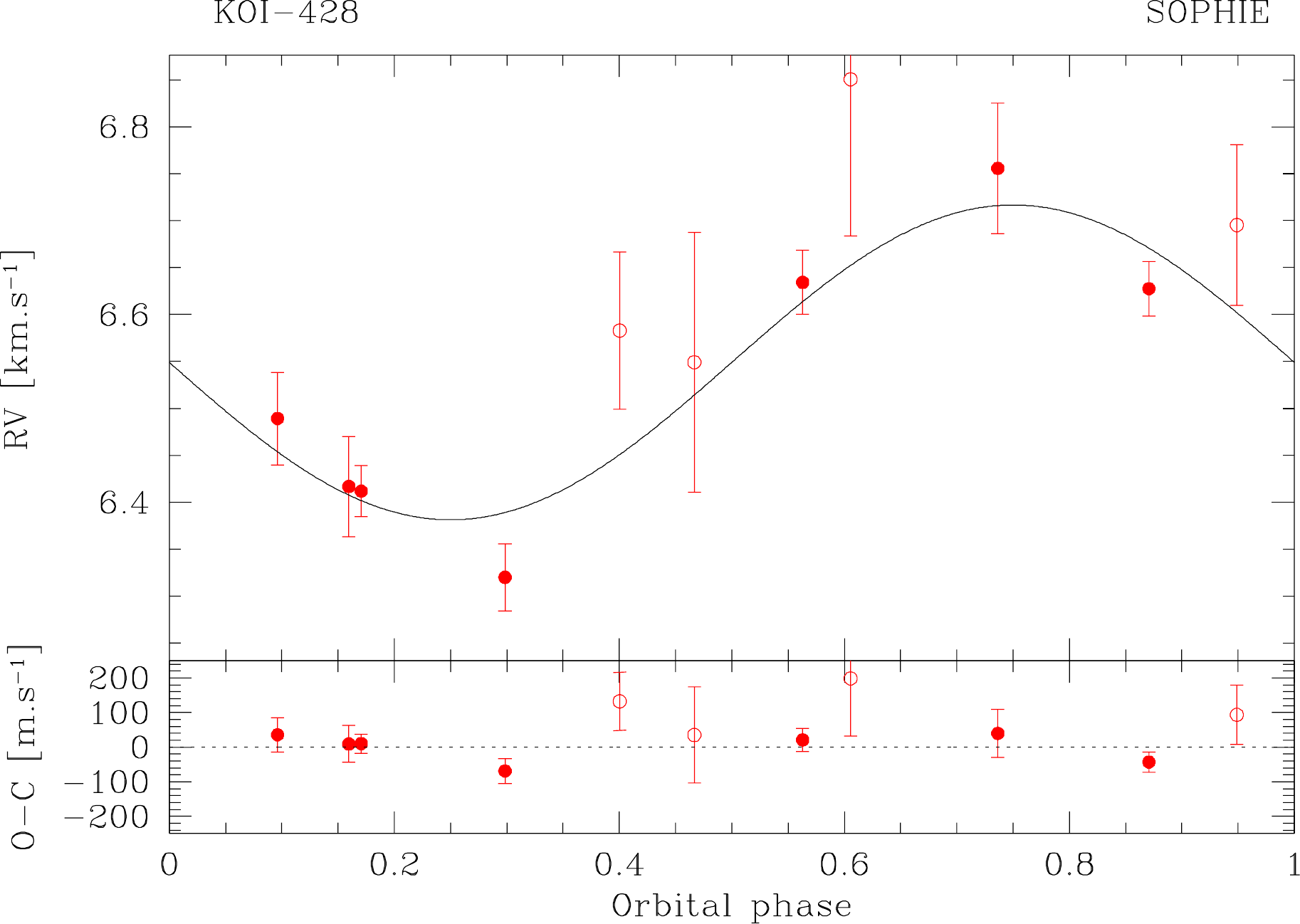}
\caption{{Phase-folded radial velocity curve with our best-fit (top panel) and residuals (bottom panel). Open circles represent the four observations strongly affected by the scattered Moon light.}}
\label{428ph}
\end{center}
%\vspace{-0.5cm}
\end{figure}

To assess the possibility that the RV variations are due to a blended binary, we checked both the variations of the bisector span of the CCF as well as the dependences of the RV variations with different cross-correlation masks \citep{2009IAUS..253..129B}. The CCFs were computed with a F0V, G2V and K5V masks for all the spectra and no significant differences were found in the amplitude of the RV variations ($K_{F0V}=177 \pm 44 \ms$, $K_{G2V}=179 \pm 27\ms$, $K_{K5V}=194 \pm 32 \ms$). The bisector span, listed in Table \ref{428rv} and plotted in Fig. \ref{428bis} does not reveal any significant variations at a level more than two times smaller than the RV changes. Moreover, neither radial velocities, nor their residuals present any significant correlation with the bisector (We find a Spearman's rank-order correlation coefficient between RV and bissector of $-0.17 \pm 0.31$). These tests allow us to secure the planetary nature of KOI-428b.

\begin{figure}[]
\begin{center}
\includegraphics[width=8.5cm]{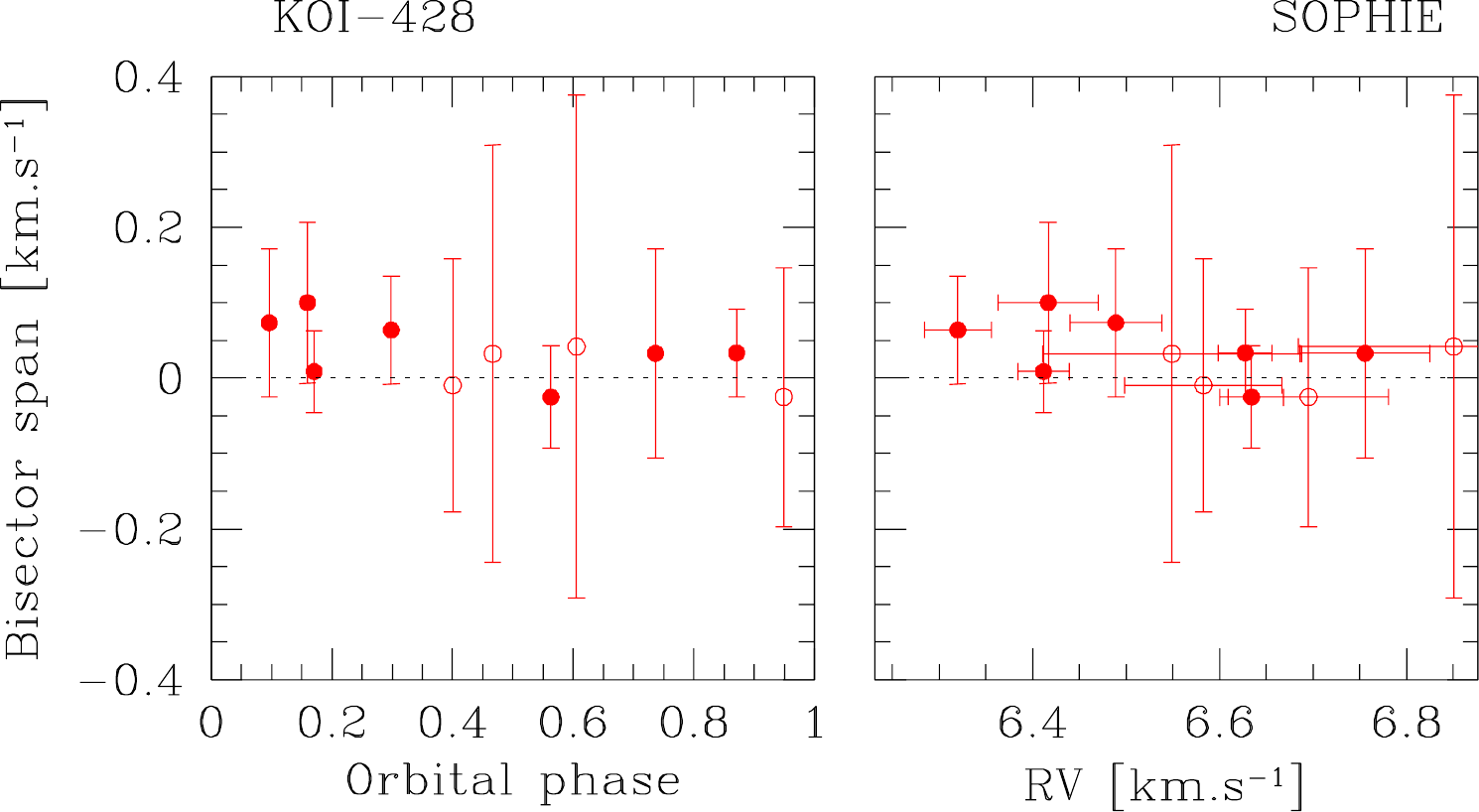}
\caption{Bisector variations as function of the orbital phase (left panel) and as function of radial velocity (right panel). Open circles display the four observations affected by the background Moon light. Bisector error bars are taken as twice the radial velocity uncertainties. No significant slope is visible between radial velocity and bisector, discarding a blend scenario.}
\label{428bis}
%\vspace{-0.6cm}
\end{center}
\end{figure}

\section{Spectral classification}
\label{spclass}
We performed the spectroscopic analysis of the parent star KOI-428 using SOPHIE spectra. Individual spectra have a low signal-to-noise, which does not allow a proper spectral analysis. Therefore, we co-added order per order the first 5 spectra not affected by the Moon light once set to the barycentric rest frame. We finally get a co-added spectrum with a S/N $\approx$ 82 per element of resolution on the continuum at 5550~\AA\ with a spectral resolution of 39,000. We derived the star's photospheric parameters using the semi-automatic software package VWA \citep{2002A&A...389..345B, 2008A&A...478..487B} as described in \citet{2010arXiv1005.3208B}. We derived \vsini = 9.0 $\pm$ 2 \kms, $v_{micro}$ = 1.0 $\pm$ 1 \kms, \teff = 6510 $\pm$ 100 K and \met = $0.1^{_{+0.15}}_{^{-0.1}}$.  We estimated the surface gravity by fitting synthetic spectra to the wings of the \ion{Mg}{I}~b and \ion{Na}{I}~D lines adopting the abundances derived from weak lines. We found \logg = 4.1 $\pm$ 0.2, a value that the quite low S/N combined to the moderate spectral resolution did not allow us to accurately estimate. The derived stellar parameters are reported in Table \ref{starplanetparam}. Our stellar parameters are significantly different from those given by \citet{2010arXiv1006.2799B} (\teff=6127K and \logg=4.55). However these authors note that ``spectroscopic observations have not been made for the released stars, so it is important to recognize that some of the characteristics listed for the stars are uncertain, especially surface gravity." Using eq. 3 in \citet{2010A&A...512A..54C} with the J and K magnitudes from the 2MASS archive (listed in Table \ref{keplerID}) and a reddening E(J-K)=0.074 \citep[computed from][]{1989ApJ...345..245C} we have computed a photometric temperature \teff= 6604 $\pm$ 132 K, which is in good agreement with the spectroscopic value.\\ 

%\vspace{-0.4cm}

\section{System parameters}

The \textit{Kepler} light curve was modeled using the analytical expressions from \citet{2002ApJ...580L.171M} with a linear limb-darkening law. We adjusted the radius ratio $k=R_p/R_\star$, the impact parameter $b$, the system scale $a/R_\star$, and the ephemerides parameters $P$ and $T_{tr}$. The eccentricity of the orbit was set to zero in the light-curve analysis, since there is no compelling evidence from the radial velocity measurements that the orbit is eccentric. The coefficient of the limb darkening law was not adjusted, but was allowed to vary within the limits defined by the spectral analysis of Sect.~\ref{spclass} (see below). On the other hand, the coarse sampling of the \emph{Kepler} Long Cadence light curves has been shown to produce an overestimation of the impact parameter \citep{2010arXiv1004.3538K} and hence a smaller measured stellar density. To avoid this, we follow the solution proposed by \citet{2010MNRAS.408.1758K}, i.e. we generate an over sampled model light curve which we bin to match the sampling rate of the \emph{Kepler} data before comparing it to the actual measurements. For the present analysis our model light curve was five times as densely sampled as the \emph{Kepler} data.

The best-fit parameters were found by $\chi^2$-minimization using the downhill simplex algorithm \citep{neldermead65}, as implemented in the SciPy library\footnote{See the SciPy Web site at \url{http://www.scipy.org}}. The fit was repeated starting from 1500 randomly-generated starting points, drawn from uniform distributions that allowed a wide range of values for each parameter. Additionally,  the value of the limb-darkening coefficient was drawn at each iteration from a normal distribution centered on the value obtained by linear interpolation of the table by \citet{2010A&A...510A..21S} using the stellar parameters computed with VWA (see Sect.~\ref{spclass}), and standard deviation equal to half the difference between the maximum and minimum values allowed by the 1-$\sigma$ intervals of the stellar parameters. In this way, we try to take into account the systematic effects introduced by fixing this value during the fit. The best-fit model is shown in Fig.Ê\ref{428fullLC} and \ref{428transit}, and the parameters are listed in Table \ref{starplanetparam}. We also repeated the fit leaving out the first transit, which is slightly deeper than the other three (see Sect.~\ref{sect.keplerdata}). We found that $k$ is within the 1.2-$\sigma$, and the other parameters are all within 1-$\sigma$ of the values determined using the entire dataset.

To estimate the uncertainties in the obtained parameters, we constructed synthetic data sets scrambling the residuals of the best-fit model and adding them back into the model light curve. We repeated the processes 2000 times, refitting the data for each realization, and recording the obtained parameters. Here again, the starting points were randomly varied and the limb-darkening coefficient was drawn from the same distribution. If we assume that the set of transit parameters obtained this way is a good approximation of the true distribution of the best-fit parameters obtained with the downhill simplex algorithm, then we can use it to estimate the confidence intervals of these best-fit parameters. The error bars reported in Table \ref{starplanetparam} correspond to the 68\% confidence interval, defined so that the cumulative probability is 16\% above and below the upper and lower confidence limit, respectively.

Finally, the stellar density obtained from the transit fit was used to calculate the value of the surface gravity (see below), which provided a new value for $u$ from which we computed a new value for the stellar density by repeating the whole fitting process. This procedure was iterated until convergence was reached, for $u=0.548^{_{+0.015}}_{^{-0.010}}$ and the values reported in Table \ref{starplanetparam}.

\begin{figure}[]
\begin{center}
\includegraphics[width=9cm]{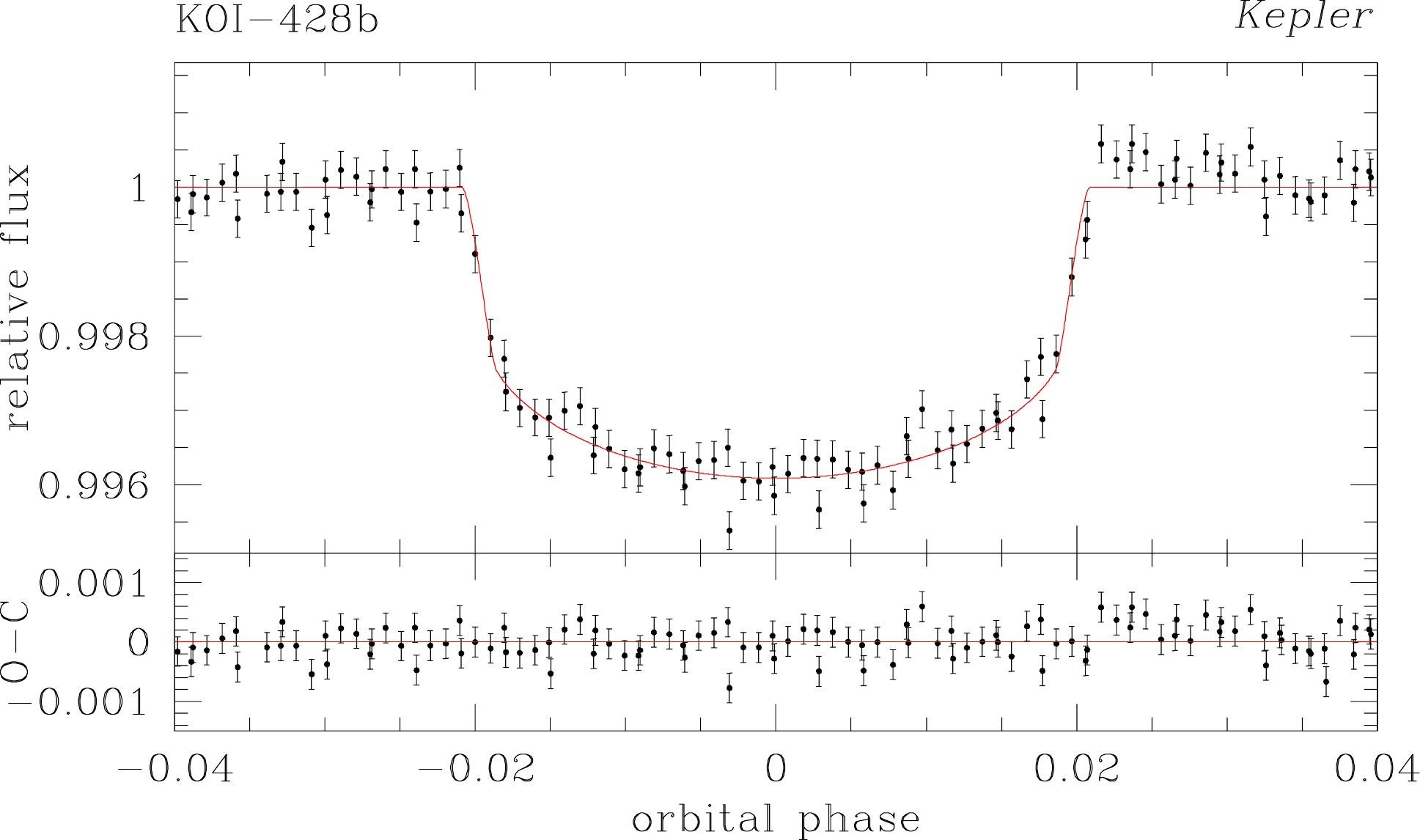}
\caption{Unbinned phase-folded light curve and residuals of the four transits of KOI-428 showing the best fit model of the transit.}
\label{428transit}
%\vspace{-0.5cm}
\end{center}
\end{figure}

We used the photospheric parameters from the spectral analysis and the stellar density from 
the transit modeling to determine the star's fundamental parameters in the (\teff, \mr) space. The location of the star in the H-R diagram was compared to  {\sl STAREVOL} evolution tracks \citep[Palacios, private com.;][]{2010ApJ...715.1539T} minimizing the usual $\chi^2$ statistics, but taking into account the time, $\tau \left[\mathrm{Gy}\right]$, a given model spends in the error box:
\begin{eqnarray} \centering
\chi^{2}_{\rm model} &=& {1\over{\tau}} \left(\frac{\rm T_{eff} - T_{eff,\rm model}}{\sigma_{\rm T_{eff}}}\right)^{2}\nonumber\\
 &+& {1\over{\tau}}\left(\frac{M_\star^{1/3}/R_\star - {M_\star^{1/3}/R_\star}_{\rm model}}{\sigma_{M_\star^{1/3}/R_\star}}\right)^{2}\nonumber\\
&+&{1\over{\tau}}\left(\frac{\rm \met - \met_{\rm model}}{\sigma_{\rm \met}}\right)^{2}\nonumber
\end{eqnarray}
We found two distincts sets of solutions which correspond to different evolutionary states: either the star is in the last main-sequence evolution stage or it is in the very beginning of the post main-sequence phase. The main-sequence solutions correspond to a star with mass of $M_\star$=1.62 $\pm$ 0.10 \Msun~ which is close to the red hook at main-sequence turnoff.  A similar situation is reported for another planet host-star, Kepler-4 \citep{2010arXiv1006.2799B} whose location in the H-R diagram is also compatible with the two evolutionary phases that could not be distinguished. In the case of KOI-428, the post main sequence solutions are clearly separated from main sequence solutions and correspond to a less-massive older star $M_\star$=1.48 $\pm$ 0.06 \Msun (see Figure \ref{428tracks}).  These solutions appear as the most probable due to the rapid evolution of stars at the end of the main sequence. We therefore adopted the post main-sequence solutions. While the mass for the two evolutionary phases differ by 14\%, both sets of solutions give a radius that differs by less than 4\% and thus have nearly no impact on the size of the planet. The inferred surface gravity is \logg\ = 3.94 $\pm$ 0.32, in agreement with the spectroscopic value at 1-$\sigma$. Clearly a better estimate of the metallicity with a much higher quality spectrum would allow to clear out this uncertainty on the evolutionary status of the star and further on the mass.
The stellar properties corresponds to a spectral type F5IV. With the adopted stellar parameters, we derived for the transiting planet M$_p$ =$2.2 \pm 0.4$ \Mjup and R$_p$ = $1.17\pm 0.04 $ \Rjup.

\begin{figure}[]
\begin{center}
\includegraphics[width=9cm]{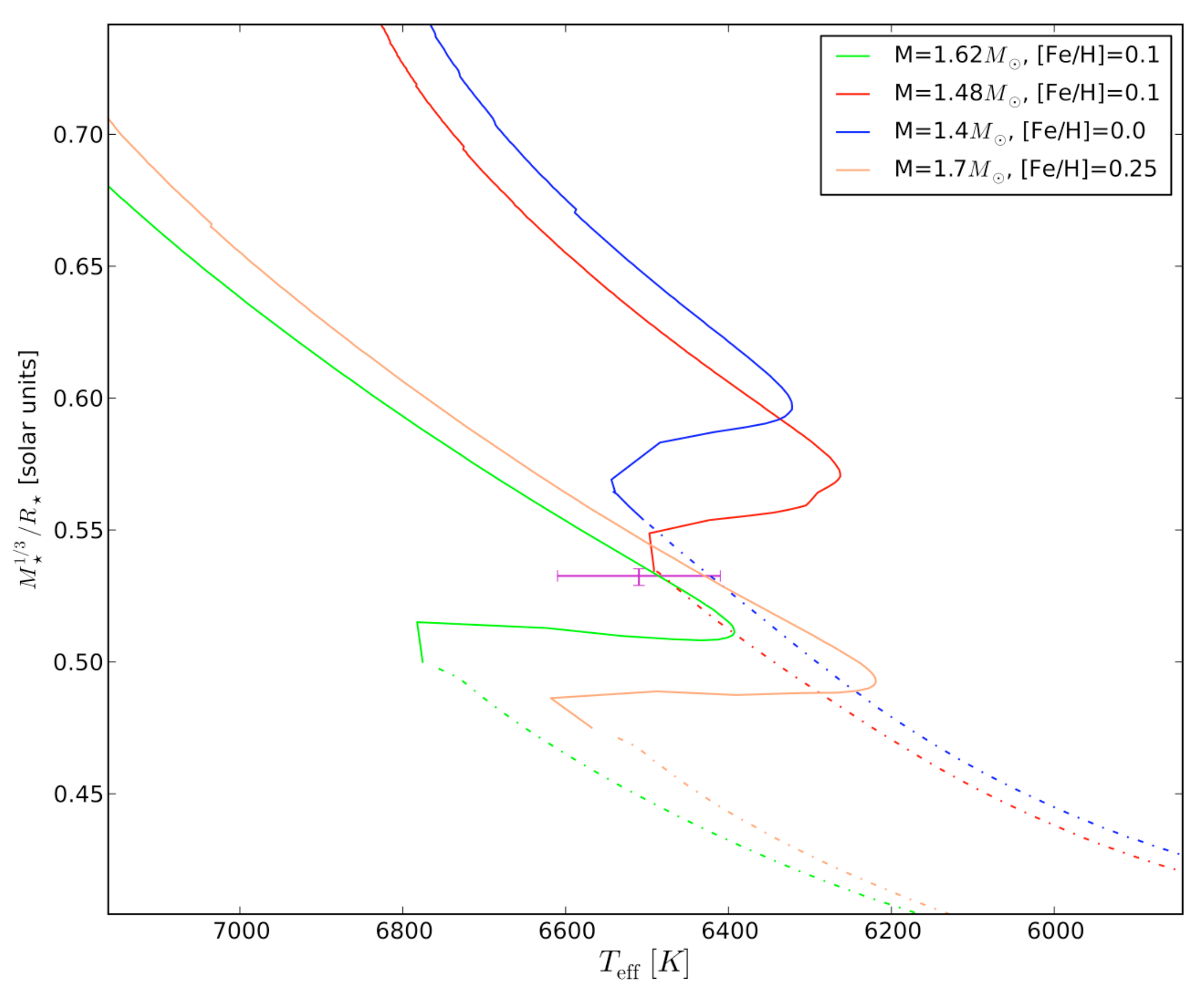}
\caption{STAREVOL evolutionary tracks models in the (\teff, \mr) space for main-sequence stars (solid lines) and post-main sequence stars (dashed-dot lines). The purple point represents the observed position of the star KOI-428 with the associated error bars in this space. The two best models are the green solid line with M$_\star=1.62\Msun$ and the adopted one in dashed-dot red line with M$_\star=1.48\Msun$ (see text).}
\label{428tracks}
%\vspace{-0.5cm}
\end{center}
\end{figure}

\begin{table}[h]
%\vspace{1cm}
\centering
\caption{Star and planet parameters.}            
%\vspace{1cm}
\begin{minipage}{9cm} 
\setlength{\tabcolsep}{1.5mm}
\renewcommand{\footnoterule}{}                          
\begin{tabular}{l l}        
\hline\hline                 
\multicolumn{2}{l}{\emph{Ephemeris}} \\
\hline
Planet orbital period $P$ [days] & $6.87349 \pm 0.00064 $  \\ 
Transit epoch $T_{tr}$ [BJD] & $2455005.5198 \pm 0.0024$  \\  
&\\
\multicolumn{2}{l}{\emph{Results from radial velocity observations}} \\
\hline    
Orbital eccentricity $e$  & 0 (fixed) \\
Semi-amplitude $K$ [\ms] & 179 $\pm$ 27 \\
Systemic velocity  $V_{r}$ [\kms] & 6.565 $\pm$ 0.020 \\
O-C residuals [\ms] & 65 \\
&\\
\multicolumn{2}{l}{\emph{Fitted transit parameters}} \\
\hline
Radius ratio $k=R_{p}/R_{\star}$ & $0.0565 \pm 0.0004$ \\ %0.05646 + 0.00039 - 0.00034 
Scaled semi-major axis $a/R_{\star}$ & $8.11\pm0.06$ \\ % 8.107 + 0.041 - 0.054
Impact parameter $b$ &  $0.045^{_{+0.011}}_{^{-0.037}}$ \\ %0.045 + 0.011 - 0.037 
&\\
\multicolumn{2}{l}{\emph{Deduced transit parameters}} \\
\hline
\mr [solar units]& $0.533 \pm 0.004 $\\ %0.5326 + 0.0027 - 0.0036
Stellar density $\rho_{\star}$ [$g\;cm^{-3}$] & $0.213 \pm 0.005$\\  
Inclination $I$ [deg] & $89.7^{_{+0.3}}_{^{-0.1}}$ \\  
Transit duration $T_{14}$ [h] & 6.86 $\pm$ 0.06 \\
&\\
\multicolumn{2}{l}{\emph{Spectroscopic parameters }} \\
\hline
Effective temperature \teff[K] & 6510 $\pm$ 100\\ 
Metallicity \met [dex]&  $0.10^{_{+0.15}}_{^{-0.10}}  $ \\   % 0.1 [-0.05  -   0.25]
Stellar rotational velocity {\vsini} [\kms] & 9.0 $\pm$ 2\\   %  9 [7  -  11]
Spectral type & F5IV\\
&\\
\multicolumn{2}{l}{\emph{Stellar and planetary physical parameters from combined analysis}} \\
\hline
Star mass $M_\star$ [\Msun] & 1.48 $\pm$ 0.06\\ 
Star radius $R_\star$[\Rsun] & 2.13  $\pm$ 0.06 \\   
Surface gravity log\,$g$$^a$ [dex]& 3.94 $\pm$ 0.32 \\ % 4.1 [3.9 - 4.3] 
Age of the star $t$ [Gy] & $2.8 \pm 0.3 $ \\
Distance of the system [pc] & 2700 $\pm$ 200 \\  
\\
Orbital semi-major axis $a^b$ [AU] & $0.080 \pm 0.003 $\\ % TBD
Planet mass $M_{p}$ [M$_J$ ] &  $2.2 \pm 0.4 $ \\ % 2.1   [Ê1.75ÊÊÊÊ-     2.49 ]
Planet radius $R_{p}$[R$_J$]  & 1.17 $\pm$ 0.04  \\ % 1.38 [ 1.19    - ÊÊÊ1,55Ê]
Planet density $\rho_{p}$ [$g\;cm^{-3}$] & $1.68^{_{+0.53}}_{^{-0.43}}$\\ 
Equilibrium temperature $^c$  $T_{eq}$ [K] &$ 1620 \pm 30 $ \\  %  1685  [ 1573    -  1796  ]
&\\       
\hline\hline
\vspace{-0.5cm}
\footnotetext[1]{derived from M$_\star$ and R$_\star$.}\footnotetext[2]{derived from a/R$_\star$ and R$_\star$.}
\footnotetext[3]{considering a zero albedo and a perfect atmospheric thermal circulation.} 
\end{tabular}
\end{minipage}
\label{starplanetparam}   
%\vspace{-0.5cm}
\end{table}

\section{Discussion and conclusion}

The hot Jupiter KOI-428b is the transiting planet with the largest host star detected so far (2.13  $\pm$ 0.06 \Rsun).
Figure \ref{rrplot} shows the radius of transiting planets as a function of the radius of their
host stars. Only 5 transiting planets orbit a star with a radius larger than 1.8 \Rsun ~ including Kepler-5 \citep{2010ApJ...713L.131K}, Kepler-7 \citep{2010ApJ...713L.140L}, TrES-4 \citep{2007ApJ...667L.195M}
and HAT-P-7 \citep{2008ApJ...680.1450P}. Transiting planets around such large stars are difficult to detect from ground-based photometry due to the long transit duration and the small transit depth. We note that only 8 candidates over the 306 published in \citet{2010arXiv1006.2799B} have an estimated host star radius larger than 1.8 \Rsun, including KOI-428. 

From our determination of the stellar radius and the projected rotational speed of the star \vsini, we note that the minimum star rotational period is $P_{rot_{min}} = 11.9 \pm 2.7 $ days, which is compatible with a synchronization at twice the planet orbital period. This is also the case of other transiting exoplanets (e.g. XO-4b \citep{2008arXiv0805.2921M}, HAT-P-6b \citep{2008ApJ...673L..79N}, HAT-P-8b \citep{2009ApJ...704.1107L}) with host-star hotter than 6000K \citep{2010A&A...512A..77L}. Additional transit measurements from \textit{Kepler} will permit to determine the true stellar rotational period (as for CoRoT-4b \citep{2008A&A...488L..43A,2008A&A...488L..47M}) and permit to understand the star-planet interaction and evolution.
%The evolve state of KOI-428 suggest that its main-sequence radius was smaller. Assuming no mass-loss and a perfect conservation of the angular momentum at the end of the main-sequence, this main-sequence radius should was ${\left(R_\star\right)}_{MS} \sim 1.4 \Rsun$ with a ${\left(\vsini\right)}_{MS} = 13.7 \pm 3.0 \kms $ which give a ${\left(P_{rot_{min}}\right)}_{MS} \sim 5.2~d$

%Additional transit measurements from \textit{Kepler} will definitively permit to reduce the transit fit uncertainties due to the sparse sampling and the systematics affecting some individual transits. 
Planetary mass error bars are dominated by radial velocity uncertainties and stellar characterization uncertainties. Additional higher S/N spectra which permit a better estimation of \met   and \teff  will definitely reduce the star mass uncertainties and the discrepancy between evolution solutions. %Furthermore, higher S/N spectra will also allow a better constraint of the system parameters, especially the stellar metallically.

This detection demonstrates the efficiency of small telescopes with dedicated instrumentation for the ground-based follow-up of space missions like \textit{Kepler} and \textit{CoRoT} \citep{2006cosp...36.3749B}. The equivalent of 1 night on a 2-m telescope led to establish the planetary nature of a transiting candidate and to determine its mass. Only the follow-up of a significant amount of \textit{Kepler} candidates will provide the real fraction of true transiting exoplanets.

\begin{figure}[h]
\begin{center}
\includegraphics[width=9cm]{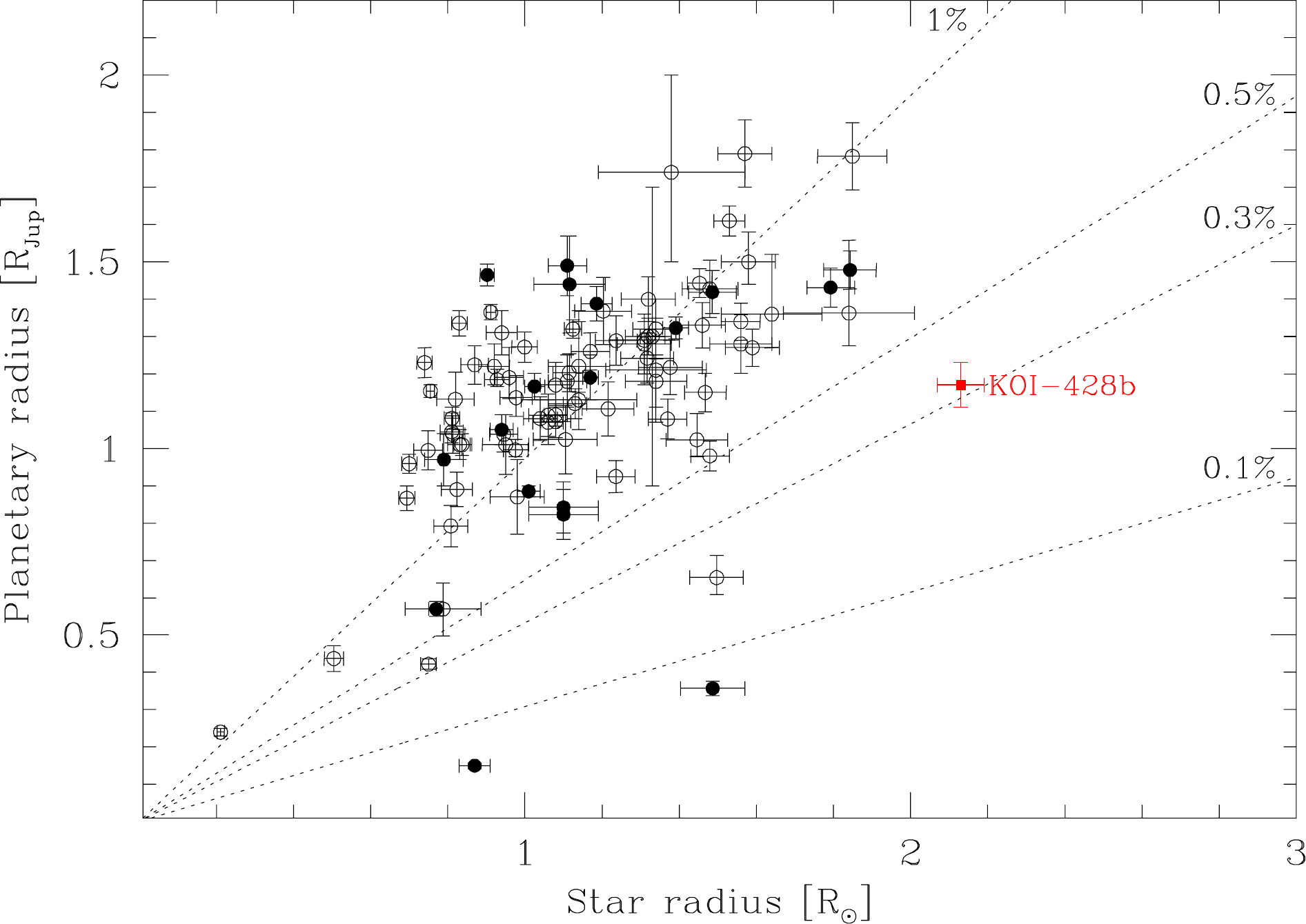}
\caption{Radius-radius diagram of all transiting exoplanets discovered so far, overplotted with curves of equal transit depth. Open circles are transiting planets discovered by ground-based photometry, filled circles are those discovered by space-based photometry and the red square is KOI-428.}
\label{rrplot}
%\vspace{-0.5cm}
\end{center}
\end{figure}

\begin{table}[h]
\centering
\setlength{\tabcolsep}{1.0mm}
\renewcommand{\footnoterule}{}                          
\begin{minipage}[t]{7.5cm} 
\caption{SOPHIE measurements of KOI-428.}
\begin{tabular}{ccccc}
\hline
\hline
BJD & RV & $\pm 1\sigma_{rv}$ & BIS & S/N/pix \\
(-2 455 000) & [\kms] & [\kms] & [\kms] & @550nm \\
\hline
392.5085           & 6.320 & 0.036 &   0.064 & 15.7\\
395.5188           & 6.755 & 0.069 &   0.033 & 10.2\\
396.4422           & 6.627 & 0.029 &   0.033 & 20.2\\
398.5034           & 6.411 & 0.027 &   0.009 & 21.0\\
400.5389$^a$  & 6.549 & 0.138 &   0.032 & 12.0\\
401.4932$^a$  & 6.850 & 0.166 &   0.042 & 10.9\\
425.4872           & 6.489 & 0.049 &   0.073 & 12.0\\
431.3489$^a$  & 6.695 & 0.086 &  -0.025 & 18.3\\
441.3261$^a$  & 6.582 & 0.084 &  -0.009 & 15.8 \\
442.4435           & 6.634 & 0.034 &  -0.025 & 15.5 \\
453.4184           & 6.417 & 0.053 &  0.100  & 16.1\\
\hline
\hline
\end{tabular}
\vspace{-0.3cm}
\footnotetext[1]{Measurement affected by scattered Moon light.}
\label{428rv}
\end{minipage}
\end{table}

\begin{acknowledgements}
We thank the technical team at the Observatoire de Haute-Provence for their support with the SOPHIE instrument and the 1.93-m telescope and in particular the essential work of the night assistants. We are grateful to the \textit{Kepler} Team for giving public access to the corrected \textit{Kepler} light curve of Q1 data set and for publishing a list of good planetary candidates to follow-up. Financial support for the SOPHIE Consortium from the ÒProgramme national de planŽtologieÓ (PNP) of CNRS/INSU, France and from the Swiss National Science Foundation (FNSRS) are gratefully acknowledged. We also acknowledge support from the French National Research Agency (ANR-08-JCJC-0102-01). A.E. is supported by a fellowship for advanced researchers from the Swiss National Science Foundation (grant PA00P2\_126150/1). D.E. is supported by the Centre National d'\'Etudes Spatiales (CNES). RFD would like to thank J.-M.~D\'esert and A.~S.~Bonomo for some fruitful discussions. AS would like to thank J.-C.~Gazzano for all interesting discussions. The authors also thank the referee for his/her fruitful comments and suggestions.
\end{acknowledgements}


\begin{thebibliography}{}
\bibitem[Aigrain et al.(2008)]{2008A&A...488L..43A} Aigrain, S., Cameron, A.~C., Ollivier, M., et al.\ 2008, \aap, 488, L43
%\bibitem[Alonso et al.(2008)]{2008A&A...482L..21A} Alonso, R., Auvergne, M., Baglin, A., et al.\ 2008, \aap, 482, L21
\bibitem[Baglin et al.(2006)]{2006cosp...36.3749B} Baglin, A., Auvergne, M., Boisnard, L., et al.\ 2006, 36th COSPAR Scientific Assembly, 36, 3749
\bibitem[Baranne et al.(1996)]{1996A&AS..119..373B} Baranne, A., Queloz, D., Mayor, M., et al.\ 1996, \aaps, 119, 373
\bibitem[Barge et al.(2008)]{2008A&A...482L..17B} Barge, P., Baglin, A., Auvergne, M., et al.\ 2008, \aap, 482, L17
%\bibitem[Boisse et al. (2010)]{HDisa}Boisse, I., Eggenberger, A., Santos, N.~C., et al., \ 2010, \aap, in press, arXiv:1006.4984 
\bibitem[Bonomo et al.(2010)]{2010arXiv1006.2949B} Bonomo, A.~S., Santerne, A., Alonso, R., et al.\ 2010, \aap, in press, arXiv:1006.2949
%\bibitem[Bord{\'e} et al.(2010)]{2010arXiv1008.0325B} Bord{\'e}, P., Bouchy, F., Deleuil, M., et al.\ 2010, arXiv:1008.0325
\bibitem[Borucki et al.(2010a)]{2010ApJ...713L.126B} Borucki, W.~J., Koch, D.~G., Brown, T.~M., et al.\ 2010, \apjl, 713, L126
\bibitem[Borucki et al.(2010b)]{2010arXiv1006.2799B} Borucki, W.~J., \& the Kepler Team 2010, arXiv:1006.2799
%\bibitem[Bouchy et al.(2008)]{2008A&A...482L..25B} Bouchy, F., Queloz, D., Deleuil, M., et al.\ 2008, \aap, 482, L25
\bibitem[Bouchy et al.(2009a)]{2009A&A...505..853B} Bouchy, F., H\'ebrard, G., Udry, S., et al.\ 2009a, \aap, 505, 853
\bibitem[Bouchy et al.(2009b)]{2009IAUS..253..129B} Bouchy, F., Moutou, C., Queloz, D., \& the CoRoT Exoplanet Science Team 2009b, IAU Symposium, 253, 129
%\bibitem[Bouchy et al.(2010)]{2010arXiv1006.2605B} Bouchy, F., Hebb, L., Skillen, I.,  et al.\ 2010, \aap, in press, arXiv:1006.2605
\bibitem[Bruntt et al.(2002)]{2002A&A...389..345B} Bruntt, H., Catala, C., Garrido, R., et al.\ 2002, \aap, 389, 345
\bibitem[Bruntt et al.(2008)]{2008A&A...478..487B} Bruntt, H., De Cat, P., \& Aerts, C.\ 2008, \aap, 478, 487
\bibitem[Bruntt et al.(2010)]{2010arXiv1005.3208B} Bruntt, H., Deleuil, M., Frindlund, M., et al.\ 2010, arXiv:1005.3208
%\bibitem[Cameron et al.(2007)]{2007MNRAS.375..951C} Cameron, A.~C., Bouchy, F., H\'ebrard, G., et al.\ 2007, \mnras, 375, 951
%\bibitem[Christian et al.(2009)]{2009MNRAS.392.1585C} Christian, D.~J., Gibson, N. P., Simpson, E. K., et al.\ 2009, \mnras, 392, 1585
%\bibitem[Deleuil et al.(2008)]{2008A&A...491..889D} Deleuil, M., Deeg, H.~J., Alonso, R., et al.\ 2008, \aap, 491, 889
\bibitem[Cardelli et al.(1989)]{1989ApJ...345..245C} Cardelli, J.~A., Clayton, G.~C., \& Mathis, J.~S.\ 1989, \apj, 345, 245 
\bibitem[Casagrande et al.(2010)]{2010A&A...512A..54C} Casagrande, L., Ram{\'{\i}}rez, I., Mel{\'e}ndez, J., Bessell, M., \& Asplund, M.\ 2010, \aap, 512, A54 
\bibitem[Dunham et al.(2010)]{2010ApJ...713L.136D} Dunham, E.~W., Borucki, W.~J., Koch, D.~G., et al.\ 2010, \apjl, 713, L136
%\bibitem[Fridlund et al.(2010)]{2010A&A...512A..14F} Fridlund, M., H\'ebrard, G., Alonso, R., et al.\ 2010, \aap, 512, A14
%\bibitem[Gandolfi et al.(subm.)]{Gandolfietal} Gandolfi, D., H\'ebrard, G., Alonso, R., et al. submitted to \aap
\bibitem[Gillon et al.(2007)]{2007A&A...471L..51G} Gillon, M., Demory, B.~O., Barman, T., et al.\ 2007, \aap, 471, L51
\bibitem[Jenkins et al.(2010a)]{2010arXiv1001.0416J} Jenkins, J.~M., Borucki, W.~J., Koch, D.~G., et al.\ 2010, arXiv:1001.0416
\bibitem[Jenkins et al.(2010b)]{2010ApJ...713L..87J} Jenkins, J.~M., Caldwell, D.~A.; Chandrasekaran, H., et al.\ 2010, \apjl, 713, L87
%\bibitem[Joshi et al.(2009)]{2009MNRAS.392.1532J} Joshi, Y.~C., Pollacco, D., Cameron, A.~C., et al.\ 2009, \mnras, 392, 1532
%\bibitem[Hebb et al.(2009)]{2009ApJ...693.1920H} Hebb, L., Cameron, A.~C., Loeillet, B., et al.\ 2009, \apj, 693, 1920
\bibitem[H{\'e}brard et al.(2008)]{2008A&A...488..763H} H{\'e}brard, G., Bouchy, F., Pont, F., et al.\ 2008, \aap, 488, 763
\bibitem[Holman et al. (2010)]{holmanetal} Holman, M.~J., Fabrycky, D.~C., Ragozzine, D., et al \ 2010, science, in press 
%\bibitem[Holman et al.(2010)]{Holmanetal}Holman, M.~J., Fabrycky, D.~C., Ragozzine, D., et al. 2010,ÊScience, in press
\bibitem[Kipping \& Bakos(2010)]{2010arXiv1004.3538K} Kipping, D.~M., \& Bakos, G.~{\'A}.\ 2010, arXiv:1004.3538 
\bibitem[Kipping(2010)]{2010MNRAS.408.1758K} Kipping, D.~M.\ 2010, \mnras, 408, 1758
\bibitem[Koch et al.(2010)]{2010ApJ...713L.131K} Koch, D.~G., Borucki, W.~J., Rowe, J.~F., et al.\ 2010, \apjl, 713, L131
%\bibitem[Lanza et al.(2009)]{2009A&A...506..255L} Lanza, A.~F., Aigrain, S., Messina, S., et al.\ 2009, \aap, 506, 255 
\bibitem[Lanza(2010)]{2010A&A...512A..77L} Lanza, A.~F.\ 2010, \aap, 512, A77 
\bibitem[Latham et al.(2009)]{2009ApJ...704.1107L} Latham, D.~W., Bakos, G. \'A., Torres, G.,et al.\ 2009, \apj, 704, 1107 
\bibitem[Latham et al.(2010)]{2010ApJ...713L.140L} Latham, D.~W., Borucki, W.~J., Koch, D.~G., et al.\ 2010, \apjl, 713, L140
%\bibitem[L{\'e}ger et al.(2009)]{2009A&A...506..287L} L{\'e}ger, A., Rouan, D., Schneider, J., et al.\ 2009, \aap, 506, 287
%\bibitem[Loeillet et al.(2008)]{2008A&A...481..529L} Loeillet, B., Shporer, A., Bouchy, F., et al.\ 2008, \aap, 481, 529
\bibitem[Nelder \& Mead(1965)]{neldermead65}Nelder, J.~A., \& Mead, R., Computer Journal , Vol. 7 (1965) , p. 308-313. 
\bibitem[Mandel \& Agol(2002)]{2002ApJ...580L.171M} Mandel, K., \& Agol, E.\ 2002, \apjl, 580, L171
\bibitem[Mandushev et al.(2007)]{2007ApJ...667L.195M} Mandushev, G., O'Donovan, F.~T.; Charbonneau, D., et al.\ 2007, \apjl, 667, L195
\bibitem[Moutou et al.(2008)]{2008A&A...488L..47M} Moutou, C., Bruntt, H., Guillot, T., et al.\ 2008, \aap, 488, L47
\bibitem[McCullough et al.(2008)]{2008arXiv0805.2921M} McCullough, P.~R., Burke, C.~J., Valenti, J.~A., et al.\ 2008, arXiv:0805.2921
\bibitem[Noyes et al.(2008)]{2008ApJ...673L..79N} Noyes, R.~W., Bakos, G. \'A.; Torres, G., et al.\ 2008, \apjl, 673, L79
\bibitem[P{\'a}l et al.(2008)]{2008ApJ...680.1450P} P{\'a}l, A., Bakos, G.~\'A., Torres, G., et al.\ 2008, \apj, 680, 1450
\bibitem[Pepe et al.(2002)]{2002A&A...388..632P} Pepe, F., Mayor, M., Galland, et al. \ 2002, \aap, 388, 632
%\bibitem[Queloz et al.(2009)]{2009A&A...506..303Q} Queloz, D., Bouchy, F., Moutou, C., et al.\ 2009, \aap, 506, 303
\bibitem[Perruchot et al.(2008)]{2008SPIE.7014E..17P} Perruchot, S., Kohler, D., Bouchy, F., et al.\ 2008, \procspie, 7014
\bibitem[Pollacco et al.(2008)]{2008MNRAS.385.1576P} Pollacco, D., Skillen, I., Cameron, A.~C., et al.\ 2008, \mnras, 385, 1576
\bibitem[Sing(2010)]{2010A&A...510A..21S} Sing, D.~K.\ 2010, \aap, 510, A21 
\bibitem[Turck-Chi{\`e}ze et al.(2010)]{2010ApJ...715.1539T} Turck-Chi{\`e}ze, S., Palacios, A., Marques, J.~P., \& Nghiem, P.~A.~P.\ 2010, \apj, 715, 1539
%\bibitem[Rauer et al.(2009)]{2009A&A...506..281R} Rauer, H., Queloz, D., Csizmadia, Sz., et al.\ 2009, \aap, 506, 281
%\bibitem[Rowe et al.(2006)]{2006ApJ...646.1241R} Rowe, J.~F., Matthews, J.~M., Seager, S., et al.\ 2006, \apj, 646, 1241
%\bibitem[Santos et al.(2002)]{2002A&A...392..215S} Santos, N.~C., Mayor, M., Naef, D., et al.\ 2002, \aap, 392, 215
%\bibitem[Shporer et al.(2009)]{2009ApJ...690.1393S} Shporer, A., Bakos, G., Bouchy, F., et al.\ 2009, \apj, 690, 1393
%\bibitem[Skillen et al.(2009)]{2009A&A...502..391S} Skillen, I., Pollacco, D., Cameron, A.~C., et al.\ 2009, \aap, 502, 391
%\bibitem[West et al.(2009)]{2009A&A...502..395W} West, R.~G., Cameron, A.~C., Hebb, L., et al.\ 2009, \aap, 502, 395
%
%

\end{thebibliography}
\end{document}